\documentclass{SciPost}

\hypersetup{
    colorlinks,
    linkcolor={red!50!black},
    citecolor={blue!50!black},
    urlcolor={blue!80!black}
}

\usepackage{hyperref}
\usepackage{float}
\usepackage[bitstream-charter]{mathdesign}
\DeclareSymbolFont{usualmathcal}{OMS}{cmsy}{m}{n}
\DeclareSymbolFontAlphabet{\mathcal}{usualmathcal}

\fancypagestyle{SPstyle}{
\fancyhf{}
\lhead{\colorbox{scipostblue}{\bf \color{white} ~SciPost Physics Proceedings }}
\rhead{{\bf \color{scipostdeepblue} ~Submission }}

\fancyfoot[C]{\textbf{\thepage}}
}

\begin{document}

\pagestyle{SPstyle}

\begin{center}{\Large \textbf{\color{scipostdeepblue}{
Recent measurements of top-quark cross sections at CMS\\
}}}\end{center}

\begin{center}\textbf{
Javier del Riego\textsuperscript{1$\star$} on behalf of the CMS Collaboration
}\end{center}

\begin{center}
{\bf 1} Universidad de Oviedo, Spain
\\
$\star$ \href{mailto:email1}{\small javier.del.riego.badas@cern.ch}\
\end{center}

\definecolor{palegray}{gray}{0.95}
\begin{center}
\colorbox{palegray}{
  \begin{tabular}{rr}
  \begin{minipage}{0.36\textwidth}
    \includegraphics[width=60mm,height=1.5cm]{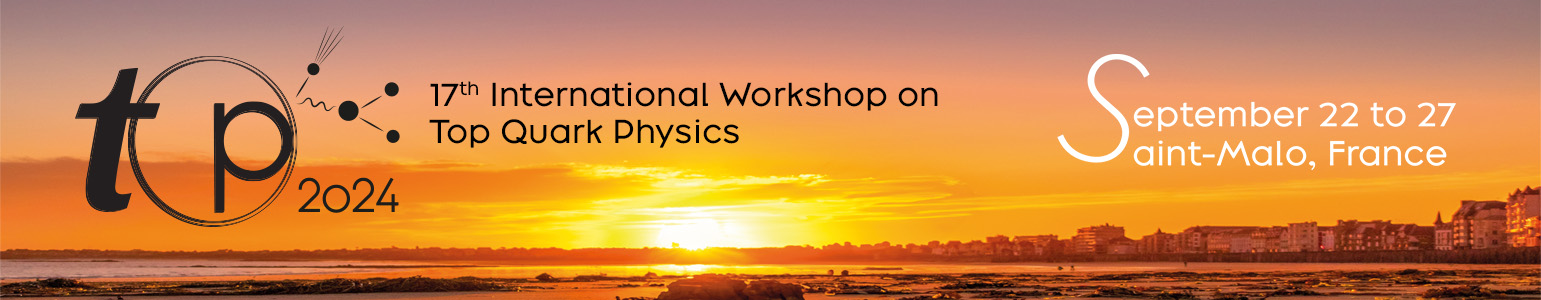}
  \end{minipage}
  &
  \begin{minipage}{0.55\textwidth}
    \begin{center} \hspace{5pt}
    {\it The 17th International Workshop on\\ Top Quark Physics (TOP2024)} \\
    {\it Saint-Malo, France, 22-27 September 2024
    }
    \doi{10.21468/SciPostPhysProc.?}\\
    \end{center}
  \end{minipage}
\end{tabular}
}
\end{center}

\section*{\color{scipostdeepblue}{Abstract}}
\textbf{\boldmath{%
This contribution presents recent cross section measurements for top quark pair production ($\mathrm{t\bar{t}}$) and single top quark production in association with a W boson (tW) in proton-proton collisions with the CMS experiment at LHC. For $\mathrm{t\bar{t}}$ production, data from 2017 at a center-of-mass (CM) energy of 5.02 TeV with an integrated luminosity of 302 pb$^{-1}$ were analyzed, with a measured inclusive cross section of $62.3 \pm 1.5 \mathrm{(stat)} \pm 2.4 \mathrm{(syst)}\allowbreak \pm 1.2 \mathrm{(lumi)}$ pb, consistent with Standard Model (SM) predictions. The tW measurement was performed using data from 2022 with an integrated luminosity of 34.7 fb$^{-1}$ at a CM energy of 13.6 TeV, yielding a cross section of $82.3 \pm 2.1 \mathrm{(stat)} ^{+9.9}_{-9.7} \mathrm{(syst)} \pm 3.3 \mathrm{(lumi)}$ pb, in agreement with SM predictions. It includes also differential cross sections which are in good agreement with next-to-leading order in perturbative quantum chromodynamics.
}}

\vspace{\baselineskip}

\noindent\textcolor{white!90!black}{%
\fbox{\parbox{0.975\linewidth}{%
\textcolor{white!40!black}{\begin{tabular}{lr}%
  \begin{minipage}{0.6\textwidth}%
    {\small Copyright attribution to authors. \newline
    This work is a submission to SciPost Phys. Proc. \newline
    License information to appear upon publication. \newline
    Publication information to appear upon publication.}
  \end{minipage} & \begin{minipage}{0.4\textwidth}
    {\small Received Date \newline Accepted Date \newline Published Date}%
  \end{minipage}
\end{tabular}}
}}
}


\vspace{10pt}
\noindent\rule{\textwidth}{1pt}
\tableofcontents
\noindent\rule{\textwidth}{1pt}
\vspace{10pt}


\section{Introduction}
\label{sec:intro}

The top quark is the heaviest known fundamental particle and plays a critical role in particle physics. Its unique properties, such as its mass or its short lifetime, 
make it an ideal subject for studying fundamental aspects of the Standard Model (SM). More precisely, the top quark pair ($\mathrm{t\bar{t}}$) production is of crucial importance for precision tests of SM predictions and serves as a window to beyond-the-SM (BSM) physics that could modify its cross section. On the side of the production of a single top quark in association with an on-shell W boson (tW), 
it is sensitive to the $V_{tb}$ parameter of the Cabbibo-Kobayashi-Maskawa matrix and also to BSM physics\cite{Tait:2000sh}. Moreover, it is interesting for its interference with $\mathrm{t\bar{t}}$ production \cite{White:2009yt} and due to its role as important background in other analyses.

This summary covers two recent analyses by the CMS Collaboration\cite{Chatrchyan:2008zzk,cmsrun3}: first, $\mathrm{t\bar{t}}$ production at a center-of-mass (CM) energy of 5.02 TeV and integrated luminosity of 302 pb$^{-1}$, which is the most precise CMS cross section measurement of the process at that CM energy, in a clean scenario of low pileup conditions and low energy spectrum. Also, tW  production at a CM energy of 13.6 TeV, utilizing an integrated luminosity of 34.7 fb$^{-1}$ from the 2022 LHC run is covered. This constituted the first single-top quark LHC result from Run 3 and includes both inclusive and differential cross section measurements, whereas the $\mathrm{t\bar{t}}$ analysis includes just the inclusive measurement. This contribution is based on the publicly available CMS analyses \cite{ttbarpas,tw}\footnote{The $\mathrm{t\bar{t}}$ analysis' results are shown as presented in the conference, but have since been superseded by the results in the submitted paper\cite{ttbar}.}.

\section{Analysis of $\mathrm{t\bar{t}}$ production cross section at 5.02 TeV}
\label{sec:ttbar}

\subsection{Experimental setup and strategy}
\label{subsec:ttbarsetup}
This analysis uses data collected at 5.02 TeV with an integrated luminosity of 302 pb$^{-1}$. Further details on the dataset and Monte Carlo (MC) samples used can be found in \cite{ttbarpas}. This run benefits from a low pileup environment, 
a feature that allows for a focused investigation of $\mathrm{t\bar{t}}$ production without significant contamination from other collisions. Apart from the $\mathrm{t\bar{t}}$ signal, other processes which are considered as background are modeled using MC samples (single top $t$ channel, single top tW, a W boson in association with additional jets, known as W+jets and Drell--Yan, known as DY). Also quantum chromodynamics (QCD) multijet events are incorporated into the analysis via a data-driven estimation. In this analysis, single lepton triggers are used, thus retaining events in which the trigger detects at least one electron (muon) with $p_{\mathrm{T}}>20$ (17) GeV. 

In order to enhance 
purity, a further selection that requires events to contain exactly one lepton (either electron or muon) with $p_{\mathrm{T}}>20$ GeV, $|\eta|<2.4$, vetoing any second lepton of opposite flavor with $p_{\mathrm{T}}$ as low as 10 GeV, and at least three jets is applied. The jets are required to have $p_{\mathrm{T}}>25$ GeV and $|\eta|<2.4$, and the events are further categorized into eight categories according to the number of jets and b-tagged jets and to lepton flavor. In such a way, categories are defined for events with 3 and $\geq$ 4 jets, further divided into events with 1 or $\geq$ 2 b-tagged jets, and denoted as 3j1b, 4j1b, 3j2b and 4j2b. Finally, events are required to have a missing transverse energy greater than 30 GeV. 

In such a scenario, 4j1b, 3j2b and 4j2b categories are very pure in signal, while 3j1b has a noticeable contribution from W+jets background (and minor from other backgrounds). Therefore, in those regions a multivariate analysis (MVA) classifier (random forest) is trained to separate the $\mathrm{t\bar{t}}$ signal from the W+jets background. 

\begin{figure}[H]
\centering
\includegraphics[width=0.48\textwidth]{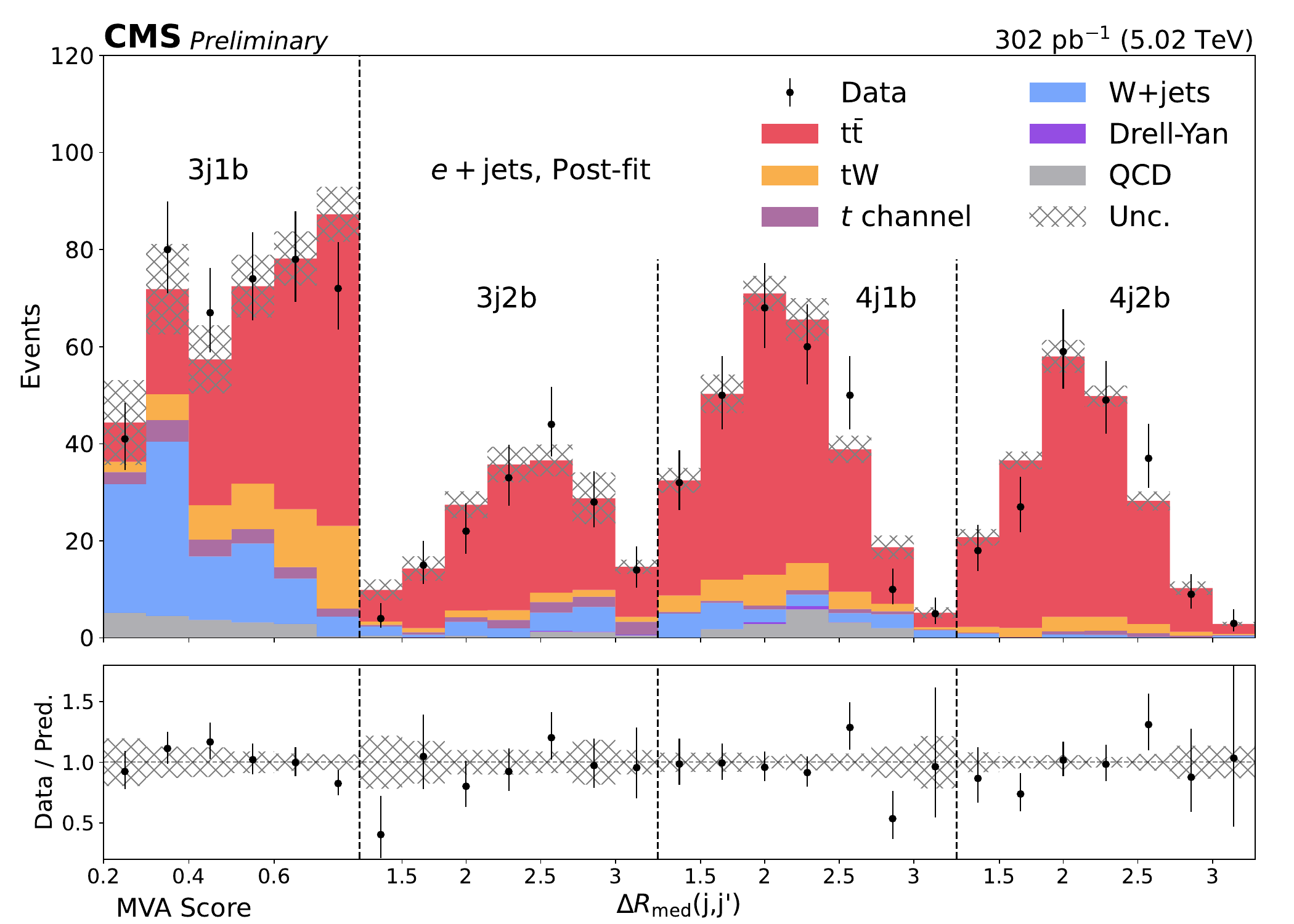}
\includegraphics[width=0.48\textwidth]{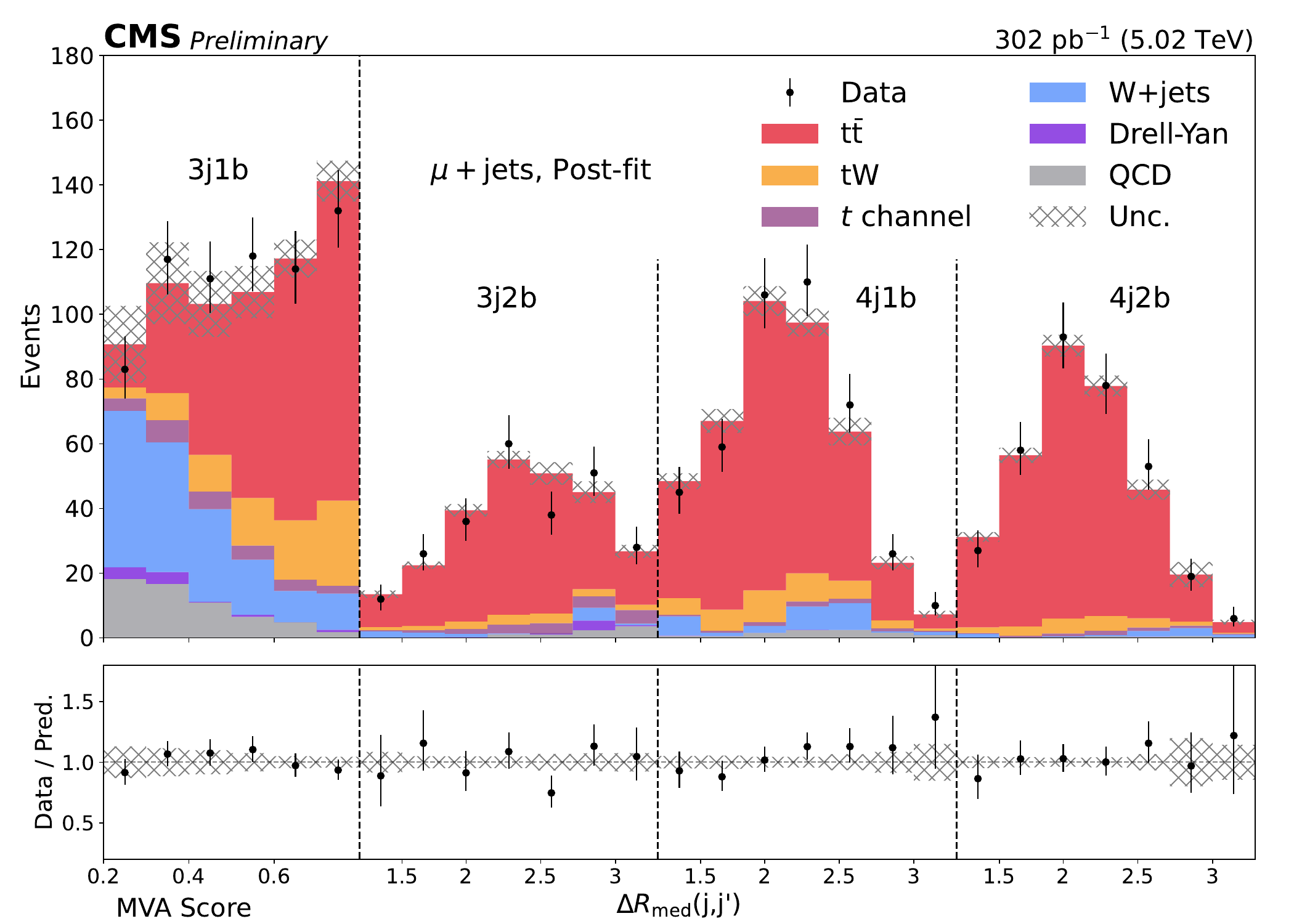}
\caption{Distributions for the electron+jets (left plot) and muon+jets (right plot) final states used for the ML fit: MVA score bins for the 3j1b category and $\Delta R_\mathrm{med}(\mathrm{j,j'})$ bins for the other categories. Post-fit data (points), predictions (filled histograms) and their ratio are shown.}
\label{fig:postfit}
\end{figure}

For the purpose of extracting the cross section, a maximum likelihood (ML) fit is performed over the 8 categories of the analysis, where the systematic uncertainties are included into the fit as nuisance parameters. In the 3j2b, 4j1b and 4j2b categories, the $\Delta R_\mathrm{med}(\mathrm{j,j'})$ distribution is used, given its discriminative power whereas in the 3j1b ones, the MVA output discriminator is the fitted variable. Those distributions (post-fit) can be seen in Figure~\ref{fig:postfit}.

\subsection{Results}
\label{subsec:ttbarresults}
\begin{figure}
\center
   \includegraphics[width=0.5\textwidth]{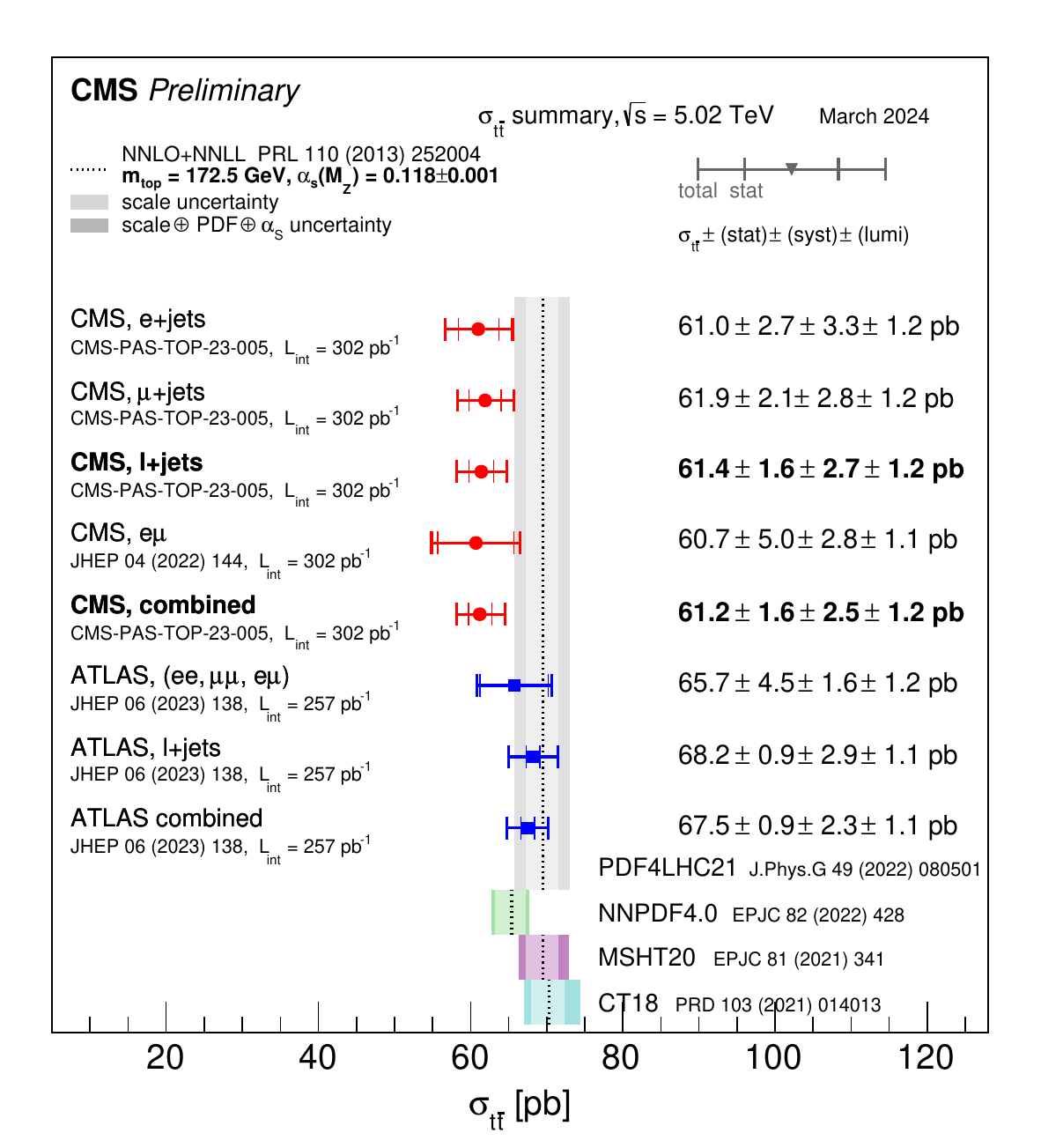}
  \caption{ Summary of the most recent $\mathrm{t\bar{t}}$ cross section measurements from the CMS and ATLAS~\cite{ATLAS:2022jbj} Collaborations, as well as theoretical predictions, for a CM energy of 5.02 TeV.}
\label{fig:summary_plot}
\end{figure}

The measured inclusive cross section after the ML fit is $\sigma(\mathrm{t\bar{t}})=62.5 \pm 1.6 \mathrm{(stat)} ^{+2.6}_{-2.5} \mathrm{(syst)} \allowbreak  \pm 1.2 \mathrm{(lumi)}$ pb. This result is combined with prior measurements in the dilepton channel\cite{combination}, resulting the combination $\sigma(\mathrm{t\bar{t}})= 62.3 \pm 1.5 \mathrm{(stat)} \pm 2.4 \mathrm{(syst)} \pm 1.2 \mathrm{(lumi)}$ pb. In both measurements, the dominant uncertainties are the ones associated with the integrated luminosity and with the b tagging scale factors for heavy flavors. Moreover, the obtained values are consistent with the SM prediction of $69.5^{+3.5}_{-3.7}$ pb at next-to-next-to-leading order (NNLO) in perturbative QCD (pQCD)\cite{tttheo}. A summary of the measurements presented here, together with previous CMS and ATLAS measurements and different theoretical predictions is shown in Figure~\ref{fig:summary_plot}.

\section{Analysis of tW production cross section at 13.6 TeV}
\label{sec:tw}

\subsection{Experimental setup and strategy}
This analysis uses data taken in 2022, corresponding to an integrated luminosity of 34.7 fb$^{-1}$, at a CM energy of 13.6 TeV. Data are selected using a combination of dilepton and single lepton triggers, requiring the events in the first case to have an electron (muon) with $p_{\mathrm{T}}>12$ GeV ($p_{\mathrm{T}}>8$ GeV) and a muon (electron) with $p_{\mathrm{T}}>23$ GeV ($p_{\mathrm{T}}>23$ GeV); and in the second, to have an electron (muon) with $p_{\mathrm{T}}>32$ (24) GeV. MC simulated samples are used to model both the signal and background contributions. The tW signal sample uses the diagram removal (DR)\cite{Frixione:2008yi} scheme to avoid double counting of Feynman diagrams in tW and $\mathrm{t\bar{t}}$ processes. Background contributions considered in this analysis are $\mathrm{t\bar{t}}$, DY, diboson processes (including WW, WZ, ZZ events and denoted as VV), vector boson production in association with a $\mathrm{t\bar{t}}$ pair (denoted as $\mathrm{t\bar{t}}V$) and W or $\mathrm{t\bar{t}}$ samples with one lepton and jets in the final state, denoted as non-W/Z. Further details on these samples can be seen in the public paper\cite{tw}.

With the purpose of enhancing the signal over background contribution, events are further selected if they contain at least a pair of opposite flavor and electric charge leptons (denoted as $e^{\pm}\mu^{\mp}$), having the leading one $p_{\mathrm{T}}>25$ GeV. In any case, all the leptons must fulfill $p_{\mathrm{T}}>20$~GeV and $|\eta|<2.4$. On top of that, events are only selected if the minimum invariant mass of all the pairs of leptons in the event is greater than 20 GeV. Events are further classified according to the exact number of jets ($n$) and b-tagged jets ($m$) in the events, satisfying all of them $p_{\mathrm{T}}>30$~GeV and $|\eta|<2.4$, and denoting each category as $n$j$m$b.

After the selection described above, 3 categories are of interest for the measurements: 1j1b (purest in signal), 2j1b and 2j2b. The three of them are used for the inclusive cross section measurement, while only 1j1b is used for the differential measurements. For the inclusive result, a similar strategy to the one described in Sec.~\ref{subsec:ttbarsetup} is followed: a ML fit to several distributions in the three mentioned categories, including the different sources of uncertainty as nuisance parameters. For the differential results, unfolding from detector to particle level is done via the \textsc{TUnfold}\cite{tunfold} technique, with a veto on events with low energy jets.

\begin{figure}[h]
\centering
\includegraphics[width=0.3\textwidth]{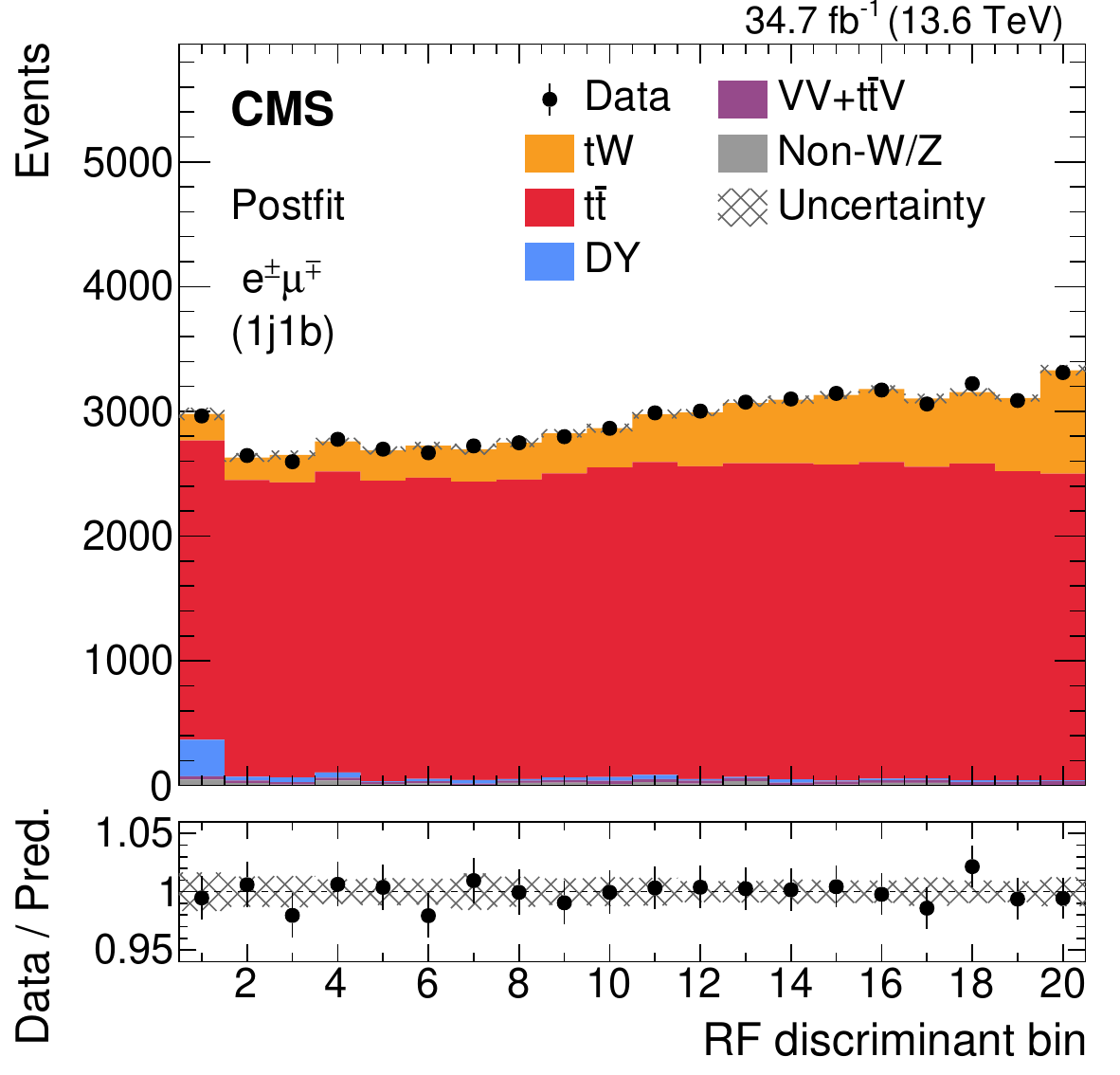}
\includegraphics[width=0.3\textwidth]{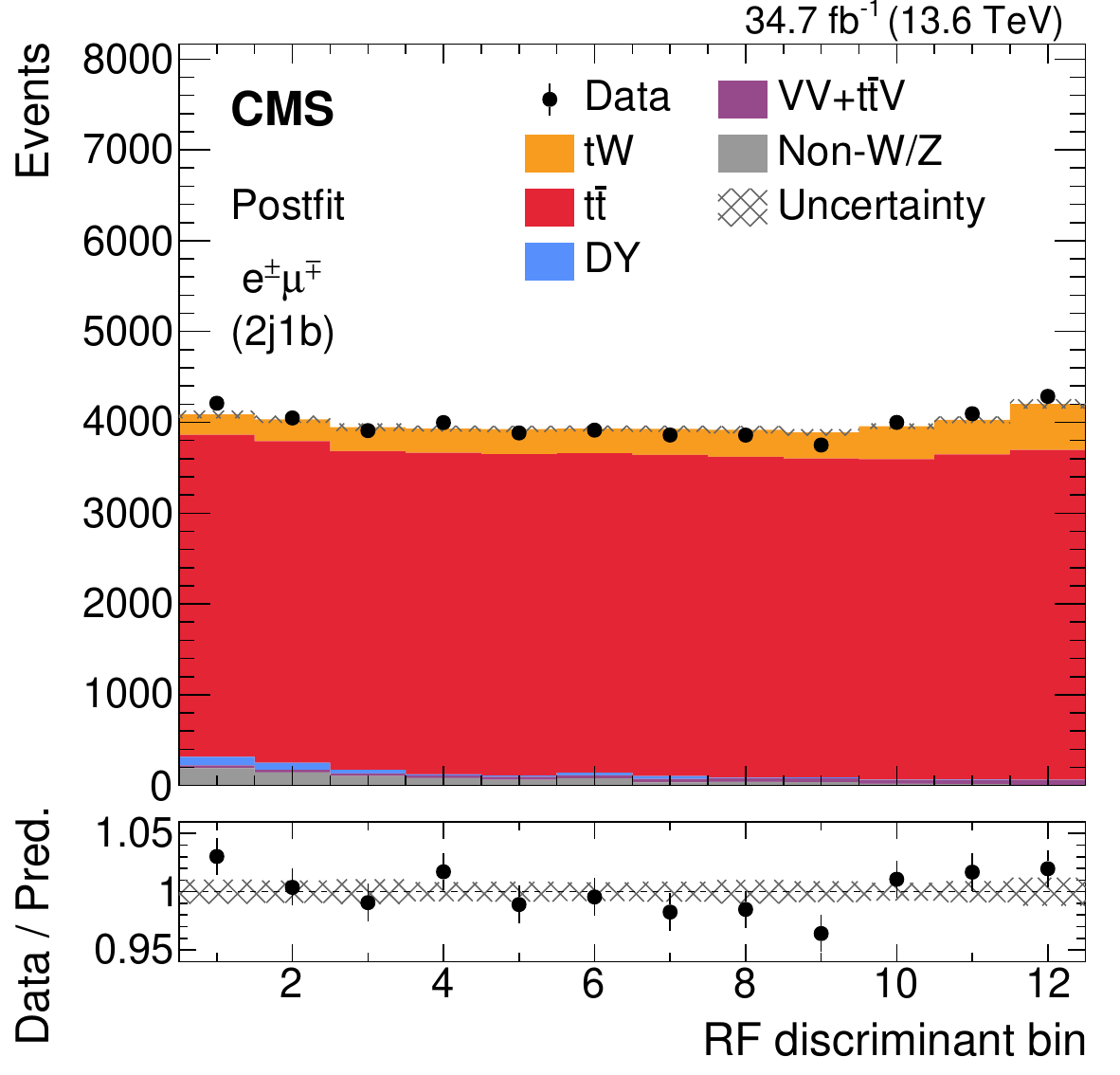}
\includegraphics[width=0.3\textwidth]{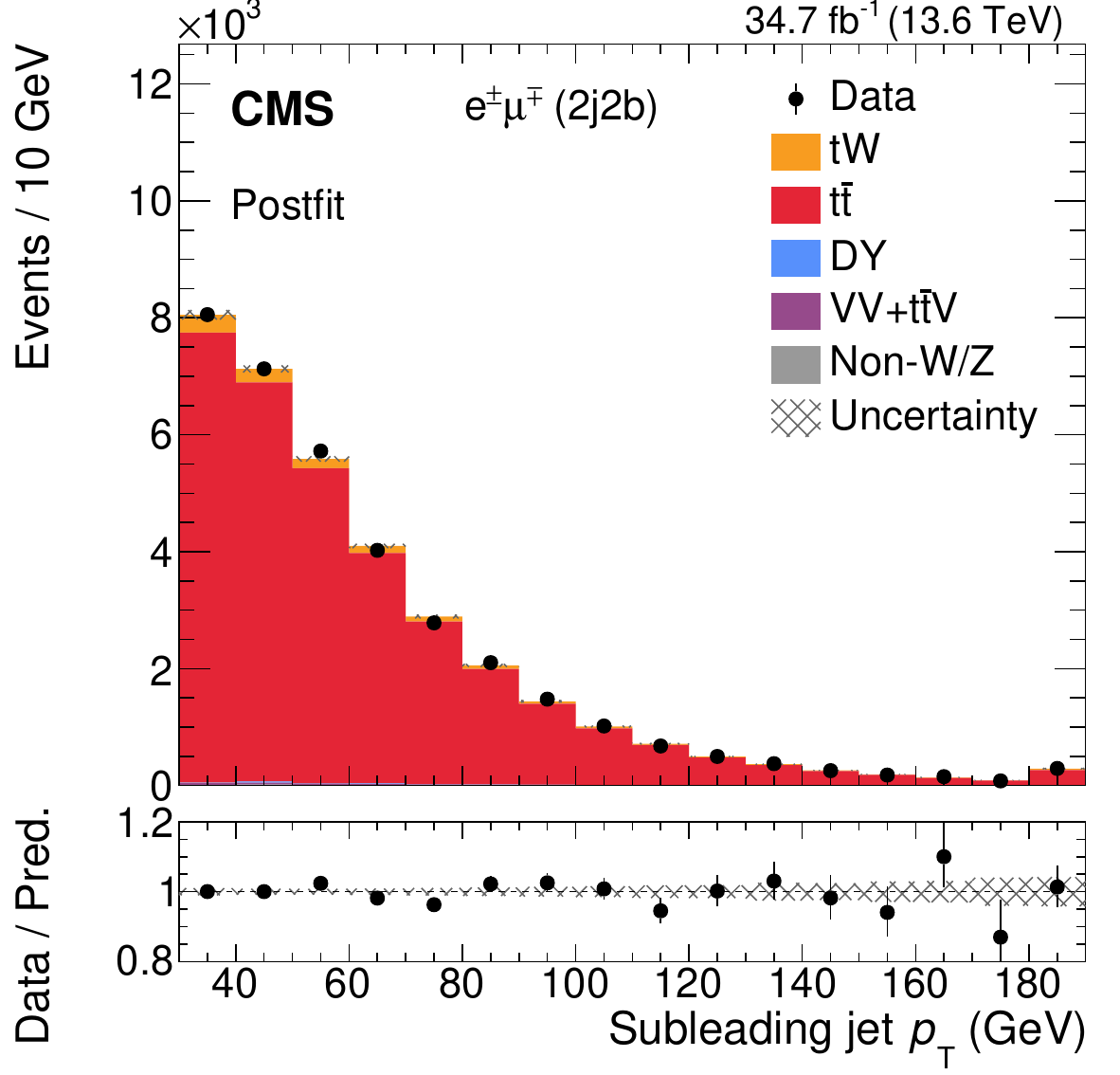}
\caption{Distributions used for the ML fit: MVA discriminant for 1j1b (left) and 2j1b (center) categories, and subleading jet $p_{\mathrm{T}}$ for 2j2b (right) category. Post-fit data (points), predictions (filled histograms) and their ratio are shown.}
\label{fig:twpostfit}
\end{figure}

\begin{figure}[hptb]
\centering
\includegraphics[width=0.3\textwidth]{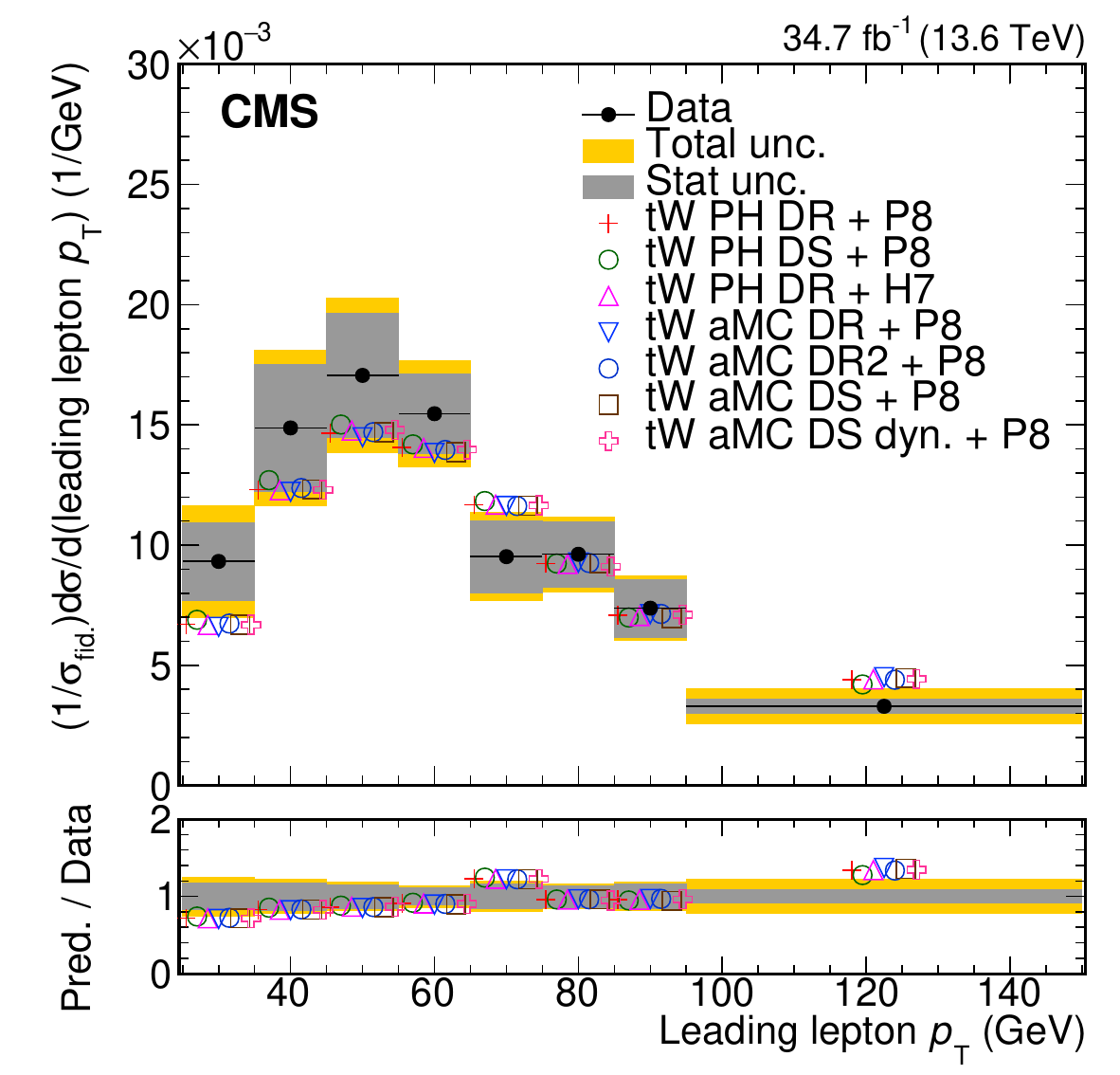}
\includegraphics[width=0.3\textwidth]{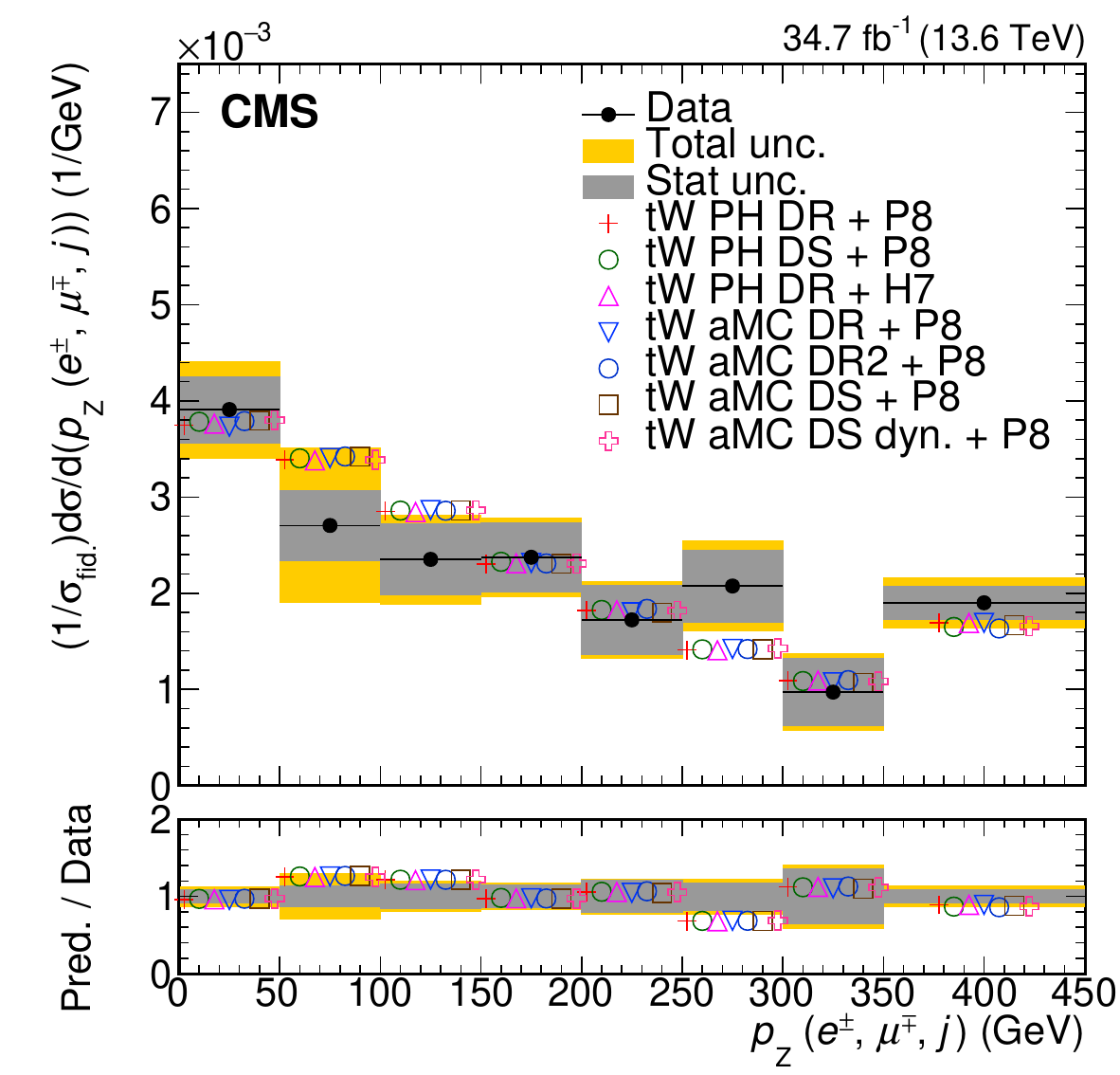}
\includegraphics[width=0.3\textwidth]{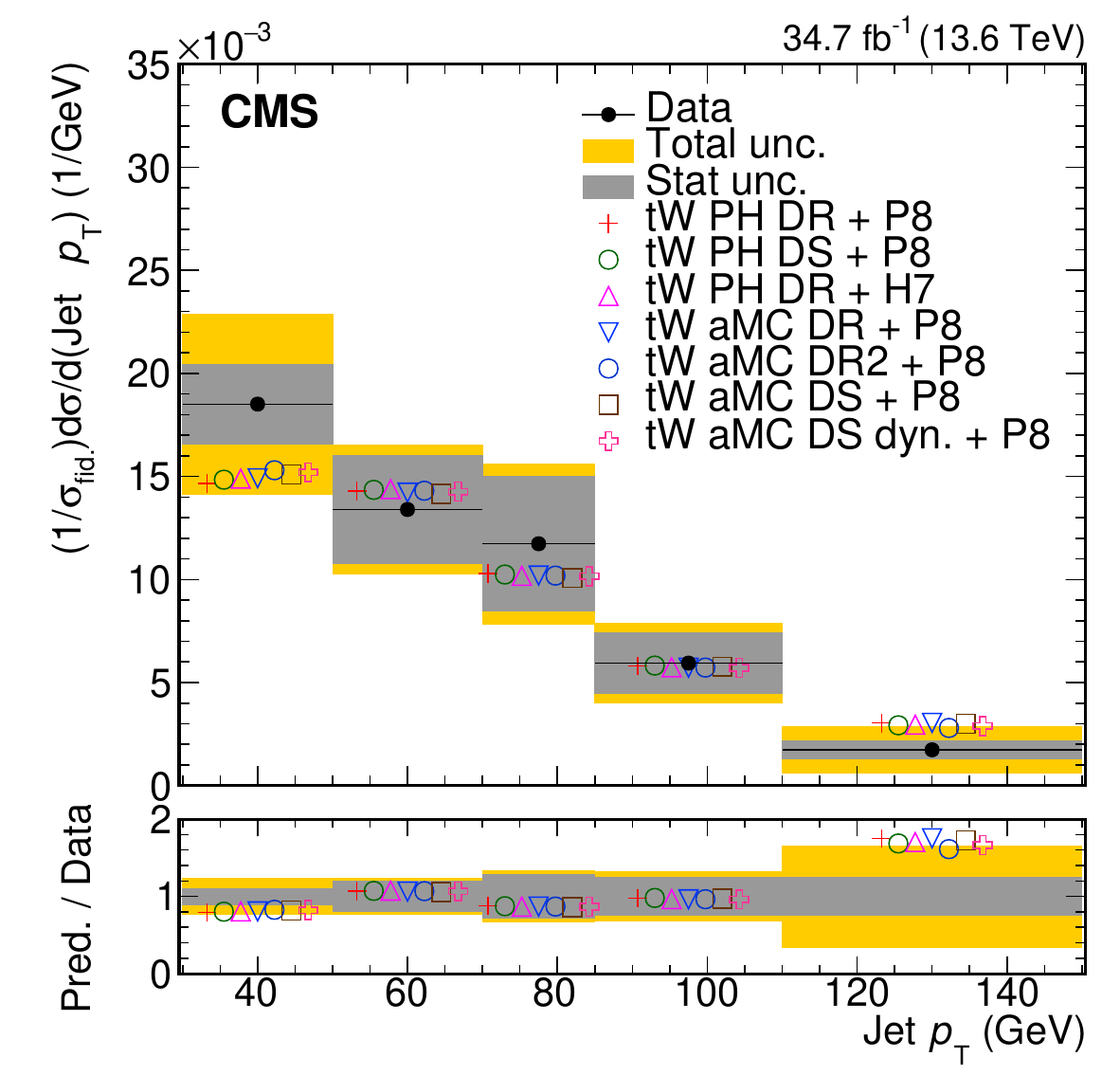} \\
\includegraphics[width=0.3\textwidth]{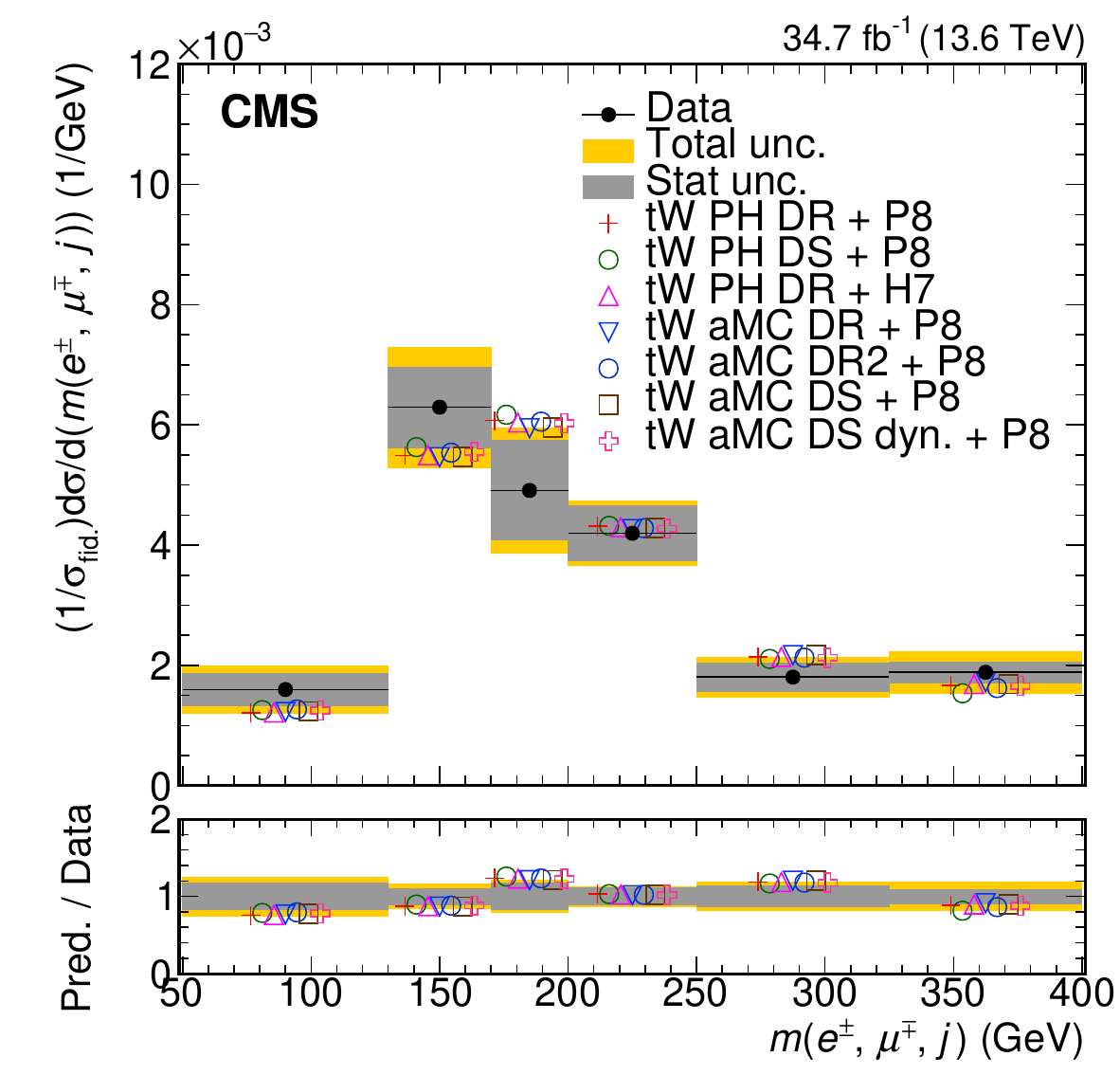}
\includegraphics[width=0.3\textwidth]{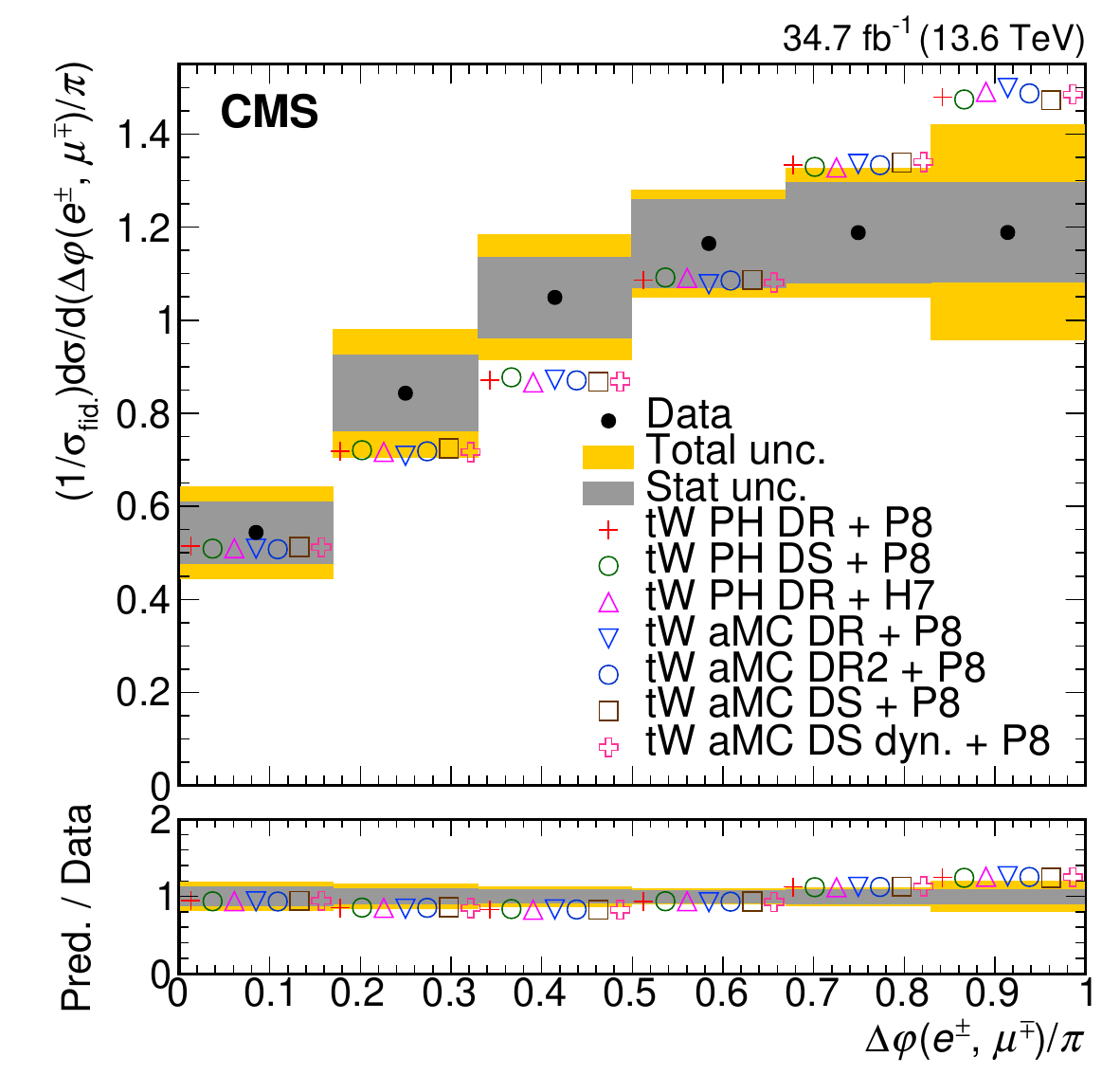}
\includegraphics[width=0.3\textwidth]{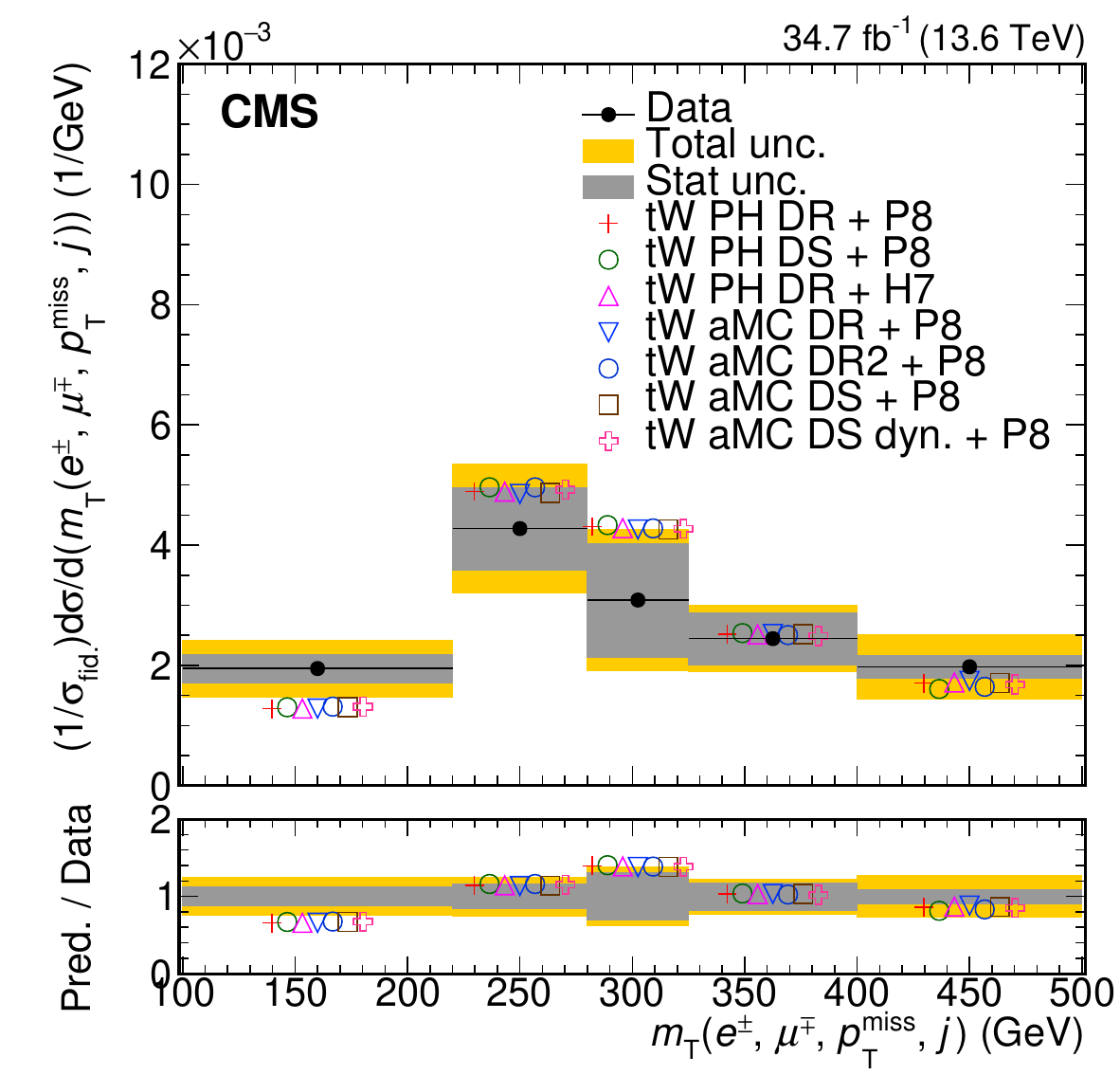}
\caption{Differential tW distributions in the 6 measured variables: leading lepton $p_{\mathrm{T}}$ (upper left), $p_{\mathrm{z}}(e^{\pm},\mu^{\pm},j)$ (upper center), jet $p_{\mathrm{T}}$ (upper right), $m(e^{\pm},\mu^{\pm},j)$ (lower left), $\Delta\phi(e^{\pm},\mu^{\pm})/\pi$ (lower center) and $m_{\mathrm{T}}(e^{\pm},\mu^{\pm},p_{\mathrm{T}}^{\mathrm{miss}},j)$ (lower right).}
\label{fig:differential}
\end{figure}

Given the challenge of separating signal from background, two independently trained MVA classifiers based on random forests are trained in the 1j1b, 2j1b categories and their discriminant distributions are used in the ML fit. In both trainings, tW is treated as signal, $\mathrm{t\bar{t}}$ as background and additionally, in the 1j1b (2j1b) category, DY (Non-W/Z) as an extra background. For the 2j2b category, the subleading jet $p_{\mathrm{T}}$ is used as input distribution into the ML fit. The three (post-fit) distributions are shown in Figure~\ref{fig:twpostfit}. In Figure~\ref{fig:differential}, the differential results, normalized to the fiducial cross section and bin width are shown for the 6 studied variables.

\subsection{Results}
The inclusive cross section after the ML fit was measured to be $\sigma(tW) = 82.3 \pm 2.1 \mathrm{(stat)} ^{+9.9}_{-9.7} \allowbreak \mathrm{(syst)} \pm 3.3 \mathrm{(lumi)}$ pb. The leading uncertainties are those related with the energy of the jets, the top quark $p_{\mathrm{T}}$ modeling and the b tagging efficiencies. This value aligns with SM prediction of $87.9\pm 3.1$ pb at approximate next-to-next-to-next-to-leading order (NNNLO) in pQCD\cite{twtheo}.

The differential cross sections were extracted for various variables and using different simulation procedures, both for the matrix element (\textsc{POWHEG, MADGRAPH5$\_$}a\textsc{MC@NLO}, with both DR and diagram substraction schemes\cite{Frixione:2008yi}) and the parton showering (\textsc{PYTHIA8, HERWIG7}) calculations. In all the variations, good agreement between data and different predictions was seen.

\section{Conclusion}
In this contribution, two recent top-associated cross section measurements performed by the CMS Collaboration are summarized. More precisely, the inclusive $\mathrm{t\bar{t}}$ cross section measurement at a CM energy of 5.02 TeV and an integrated luminosity of 302 pb$^{-1}$ and the inclusive and differential tW cross section measurements at 13.6 TeV and integrated luminosity of 34.7~fb$^{-1}$, in both cases using proton-proton collisions. All the measurements described are consistent with the SM predictions, being the former the most precise $\mathrm{t\bar{t}}$ cross section measurement performed by the CMS experiment at that CM energy, and the latter the first LHC single-top result using Run 3 data.







\bibliography{TOP2024_JavierDelRiego.bib}


\end{document}